\documentclass[12pt]{iopart}
\usepackage{graphicx}
\usepackage{iopams}
\usepackage[numbers,sort&compress]{natbib}
\usepackage{color}
\usepackage{todonotes}
\usepackage{comment}

\newcommand{\redb}[1]{\textcolor[RGB]{150,0,0}{\textbf{#1}}}

\usepackage{caption}
\begin{document}

\title{Community Detection through Vector-label Propagation Algorithms}

\author{Wenyi Fang$^{1,3,4}$, Xin Wang$^{2,3}$, Longzhao Liu$^{2,3}$, Zhaole Wu$^{2,3}$, Shaoting Tang$^{2,3,4,5}$ and Zhiming Zheng$^{2,3,4,5}$}

\address{$^1$ School of Mathematical Sciences, Peking University, Beijing, 100871, China}
\address{$^2$ School of Mathematics and Systems Science, Beihang University, Beijing 100191, China}
\address{$^3$ LMIB, NLSDE, BDBC, Beijing, 100191, China}
\address{$^4$ PengCheng Laboratory, Shenzhen, 518055 , China}
\address{$^5$ Author to whom any correspondence should be addressed}

\ead{tangshaoting@buaa.edu.cn, zzheng@pku.edu.cn}

\vspace{10pt}

\begin{abstract}
Community detection is a fundamental and important problem in network science, as community structures often reveal both topological and functional relationships between different components of the complex system. In this paper, we first propose a gradient descent framework of modularity optimization called vector-label propagation algorithm (VLPA), where a node is associated with a vector of continuous community labels instead of one label. Retaining weak structural information in vector-label, VLPA outperforms some well-known community detection methods, and particularly improves the performance in networks with weak community structures. Further, we incorporate stochastic gradient strategies into VLPA to avoid stuck in the local optima, leading to the stochastic vector-label propagation algorithm (sVLPA). We show that sVLPA performs better than Louvain Method, a widely used community detection algorithm, on both artificial benchmarks and real-world networks. Our theoretical scheme based on vector-label propagation can be directly applied to high-dimensional networks where each node has multiple features, and can also be used for optimizing other partition measures such as modularity with resolution parameters.
\end{abstract}


\section{Introduction}
The inherent community structure is ubiquitous in many natural systems and often contains abundant functional information of complex networks, such as the functions of proteins, the patterns of scientific collaboration, the word association in language evolutions and the emergence of social polarization and echo-chambers \cite{palla-2005-uncov-overl-commun, wang2020public}. Consequently, community detection is of fundamental significance for further understanding the complex interplay between network structure and dynamical processes across different fields, ranging from statistical physics, biology, ecology, economics and social science \cite{girvan-2002-commun-struc-social, fortunato-2010-commun-detec-graph, malliaros-2013-clust-commun-detec, fortunato-2016-commun-detec-networ, lima-2015-determinants}. Moreover, the process of detecting communities can even contribute to designing more effective data storage systems and improving network capacity \cite{boldi-2011-layer,cai-2018-enhan-networ-capac}.

In the field of network science, intuitively, a cluster or a community is a group of nodes within which the links are denser than the network's average level. To describe the communities in a network, we first introduce the definition of partition. Given a network $G=\left<V, E\right>$, where $V$ is the node set and $E$ is the edge set, a partition is defined as a disjoint union of $V$. Mathematically, a partition $P$ can be represented as a map from the node set $V$ to the community label set $C=\{1,2,\cdots,k\}$, such that
\begin{equation*}
\eqalign{
  P:&V \rightarrow C\\
  &i \mapsto c_i,}
\end{equation*}
where $c_i$ is the community label of node $i$, and $k$ is the number of communities. As there exist so many different partitions given a certain network, a criterion that measures the ``goodness" of the partitions is necessary. While many criteria have been proposed, such as Modularity \cite{newman-2004-fast-algor-detec}, Infomap \cite{rosvall-2008-maps-random-walks}, Surprise \cite{aldecoa-2011-decip-networ-commun} and etc, there is still no consensus on the precise definition of community structure and the best criterion for measuring partitions. Therefore, numerous community detection algorithms are developed through optimizing these metrics \cite{danon-2005-compar-commun-struc-ident}.

Another important line of research is mostly based on Label Propagation Algorithm (LPA) proposed by Raghavan et al. \cite{raghavan-2007-near-linear-time}, where the communities are divided via an `epidemic' propagation process of nodes' labels. The core idea of LPA is that the node's label should be the same as its most frequent neighbors' when the partition is optimal \cite{radicchi-2004-defining}. Initially, every node has a unique and different label. The updating rules are as follows: (1) If there is only one most common label among the neighbors of a node $i$, $i$'s label will be updated to the most common label; (2) If $i$'s neighbors have several most common labels (especially at the beginning), then $i$ randomly adopts one, with ties broken uniformly randomly; (3) LPA ends when the label of each node is in the most common label set of its neighbors. LPA has two main advantages: simplicity, and the low time complexity $O(cm)$, where $m$ is the number of edges and $c$ is the number of iterations before convergence. These characters make it easy to be applied for exploring the community structure of large-scale networks like online social networks.
	
However, LPA displays weak stability due to the processes of randomly selecting labels and breaking ties \cite{vsubelj-2011-robus-networ-commun}. In addition, for networks with weak community structures, the explosive label diffusion process can result in the emergence of a trivial and meaningless partition where most nodes hold the same label \cite{leung-2009-towar-real-time}. Therefore, many variants of LPA have been proposed to improve the performance \cite{liu-2010-advan-modul-special, leung-2009-towar-real-time,cordasco-2010-commun, vsubelj-2011-unfol-commun-large, vsubelj-2011-robus-networ-commun, li-2017-stepp-commun-detec, gui-2018-commun-discov-algor}. Leung et al. introduced a decreasing hop attenuation, by which one can restrict community sizes from being too large \cite{leung-2009-towar-real-time}. Barber and Clark devised LPAm as an optimization paradigm to maximize modularity \cite{barber-2009-detec-networ-commun}. Further, Blondel et al. proposed the famous Louvain Method which is based on the local dynamical optimization of modularity, with a first phase similar to LPAm \cite{blondel-2008-fast-unfol-commun}.

In this paper, to further overcome the shortcomings of LPA, especially to improve the performance in networks with weak community structures while maintaining LPA's low time cost, we develop a new framework of modularity optimization using gradient descent based on vector-label propagation, where the one-dimensional discrete labels in Louvain or LPAm are extended to a vector of continuous labels. Accordingly, two new community detection algorithms are proposed: vector-label propagation algorithm (VLPA) and stochastic vector-label propagation algorithm (sVLPA). We show that VLPA uncovers more topological information and optimizes modularity much better when dealing with weak community structures compared to several classic algorithms. Equipped with stochastic gradient strategies, sVLPA can further jump out of the local optimum and outperforms Louvain Methods on both artificial and real-world networks. Meanwhile, the time complexity of these two algorithms are both comparable to LPA.

The rest parts of this paper is organized as follows. First, we introduce the concept of vector-label in section \ref{vector_label} and thereby propose the detailed theoretical scheme of VLPA and sVLPA in section \ref{algorithms}. Then we discuss the choices of key parameters and show comparisons of experimental results between our new algorithms and some well-known algorithms on different benchmarks in section \ref{experiments}. Finally, the theoretical time complexity analysis is conducted in section \ref{time}, followed by conclusions and further discussions in section \ref{conclusions}.

\section{Definition of vector-labels}\label{vector_label}
As a widely used evaluation criterion for community detection, the modularity $Q$ \cite{fortunato-2006-resol-limit-commun-detec} of a given partition is defined as
\begin{equation}
Q=\frac{1}{2m}\sum_{i j}\left(A_{i j}-\frac{k_i k_j}{2m}\right)\delta (c_i, c_j),
\end{equation}
where $A_{ij}=1$ if node $i$ and $j$ are connected and $A_{ij}=0$ otherwise, $2m=\sum_{ij}A_{ij}$ is twice the number of edges, $k_i$ is the degree of node $i$, $c_i$ denotes the label of community to which node $i$ belongs, and $\delta$ is the Kronecker delta function with the following form:
\begin{equation}
\delta(x, y) = \left\{ \eqalign{
 1 \quad & x = y;\\
 0 \quad & x \not = y.
}\right.
\end{equation}
Naturally, the community detection problem can be transformed into a modularity optimization problem. To be specific, some community detection algorithms start from a trivial partition where each node has a different label, and then adjust the partition (changing community labels) via local greedy optimization in each iteration to maximize the modularity value, e.g., LPAm, LPAm+ and Louvain. While these methods are proven effective in many networks, the local greedy optimization at each round of label updating is not a guarantee of the global optimal solution in mathematics, leading to a poor performance especially when the community structure is not strong.

To this end, we propose the idea of vector-labels in pursuit of a global optimum. Inspired by some multi-label community detection algorithms \cite{gregory-2010-findin-overl-commun,lancichinetti-2011-findin-statis-signif,wu-2012-balan-multi-label}
and the extended definitions of modularity \cite{nicosia-2009-exten-defin-modul,griechisch-2011-commun-detec-by}, we relax the node's label from a positive integer to a vector $v$ such that
\begin{equation}
v\in \mathcal{U}^+(\mathcal{R}^n)=\{v\in \mathcal{R}^n|v^k\geq 0, \|v\|_2=1\},
\end{equation}
where $n$ is the number of nodes in the network, $\|\cdot\|_2$ is Euclidian 2-norm and $v^k$ is the $k$-th element of vector $v$.
As the number of communities cannot exceed the number of nodes, we determine the dimension of the vector-label space to be $n$. According to our definition, the value of $(v^k)^2$ can be considered as the probability that the node belongs to the community $k$. Therefore, when there exists an integer $k$ such that $v^k=1$, the vector-label degenerates into a conventional label.

With the above defined vector-label, modularity function $Q$ now can be extended to vector-modularity $Q^v$, which has the following form:
\begin{equation}
Q^v = \frac{1}{2m}\sum_{ij}\left(A_{ij}-\frac{k_ik_j}{2m} \right )\left<v_i,v_j\right>,
\end{equation}
where $\left<\cdot ,\cdot\right>$ is the inner product function and $v_i$ is the vector-label of node $i$. The vector-modularity $Q^v$ degenerates to the original modularity when there is only one nonzero component in each vector-label.

In general, the vectorization of labels may provide two benefits. First, vector-labels contain more information of community structures, which promotes the propagation of weak label components and helps jump out of the local optimum. Second, the introduction of $Q^v$ enables the gradient descent method to be applied, which can better optimize the modularity function.

\section{Algorithms}\label{algorithms}
Two novel community detection algorithms based on asynchronous vector-label propagation are proposed, named Vector-Label Propagation Algorithm(VLPA) and stochastic Vector-Label Propagation Algorithm(sVLPA).
Supposing that the vector-label $v_i$ is updated as $w_i$, the variation of vector-modularity can be expressed as
\begin{equation}
\Delta Q^v  = \left<w_i-v_i, \frac{1}{m}\left(\sum_j A_{ij}v_j-\sum_{j\not = i} \frac{k_ik_j}{2m}v_j\right) \right>.
\end{equation}
Since $\left<v_i,v_i\right>=1$, the second term in the inner product is the partial derivative of vector-modularity with respect to $v_i$:
\begin{equation}
\eqalign{
\frac{\partial Q^v}{\partial v_i}\nonumber&= \frac{1}{m}\left(\sum_j A_{ij} v_j-k_i\sum_{j\not=i} \frac{k_j}{2m}v_j\right) \\
&= \frac{1}{m}\left(\sum_j A_{ij} v_j+\frac{k_i^2}{2m}v_i - k_i\sum_{j} \frac{k_j}{2m}v_j\right). } \label{update}
\end{equation}
The first two terms of the last line of the equation (\ref{update}) are easy to compute because only the vector-labels of neighbours are needed, and the third term can be calculated iteratively.
For convenience, we decompose $\frac{\partial Q^v}{\partial v_i}$ as below:
\begin{equation}
\frac{\partial Q^v}{\partial v_i} = p_i + n_i,
\end{equation}
where $p_i$ is the vector containing all the positive components of $\frac{\partial Q^v}{\partial v_i}$ and $n_i=\frac{\partial Q^v}{\partial v_i}-p_i$ is the vector containing all the nonpositive components of $\frac{\partial Q^v}{\partial v_i}$.
Under the decomposition $\Delta Q^v$ is further expressed as
\begin{equation}
\eqalign{
\Delta Q^v
&=\left<w_i-v_i, \frac{\partial Q^v}{\partial v_i}\right>  \\
& = \left<w_i, \frac{\partial Q^v}{\partial v_i}\right>-\left<v_i, \frac{\partial Q^v}{\partial v_i}\right> \\
&= \left<w_i, p_i \right> + \left<w_i, n_i\right> - \left<v_i,\frac{\partial Q^v}{\partial v_i}\right>.}
\end{equation}
The third term of the last line is a constant so that only the first two terms need to be optimized in our local greedy vector-label algorithms.

According to the definition of vector-labels mentioned in Section \ref{vector_label}, each vector-label is $n$-dimensional which is time and space consuming. To reduce the time consumption and space occupation, sparse vector-labels are indispensable. In our algorithms,
each node is restricted to a limited number of communities rather than $n$ different communities.
The maximal number of communities to which a node can belong is defined as the essential dimensions of vector-labels, denoted by $d_e$.
Then, the feasible space of vector-labels can be expressed as
\begin{equation}
\mathcal{SU}^+(\mathcal{R}^n,d_e)=\{v\in \mathcal{R}^n|v^k \geq 0,\|v\|_2=1,\|v\|_0=d_e\},
\end{equation}
where $\|\cdot \|_0$ gives the number of nonzero components.
For saving the storage, every vector-label is stored as a list of pairs that each pair contains two elements: the index of the nonzero component and the corresponding value.

With the previous preparations, the critical vector-label update step in our algorithm can be summarized as finding a sparse vector-label $w \in \mathcal{SU}^+(\mathcal{R}^n,d_e)$ such that $w$ maximizes $\left<w,p\right>+\left<w,n\right>$, where $p$ and $n$ are the positive and negative part of a vector respectively.
The optimal $w$ which maximizes $\left<w,p\right>$ has the following form:
\begin{equation}
w=\sum_{k\in I}w^k e_k,
\label{optimal_w}
\end{equation}
where $I=\{i_1,i_2,\cdots,i_{d_e}\}$ satisfying $p^{i}\geq p^{j}\geq 0, \forall i \in I, \forall j\not \in I$ and $p^k$ is the $k$-th component of $p$. The proof is as followed.

Suppose that $\tilde{w}=(w^1,w^2,\cdots,w^n)\in\mathcal{SU}^+(\mathcal{R}^n,d_e)$ maximizes $\left<w,p\right>$ but dissatisfies the above formula, i.e., there exits $j_1 \not\in I$ such that $\tilde{w}^{j_1}> 0$.
According to the definition of $\tilde{w}$ (the number of nonzero components of $\tilde{w}$ is not greater than $d_e$), there exists $j_0 \in I$ such that $\tilde{w}^{j_0} = 0$.
Taking $\hat{w}=\tilde{w}-\tilde{w}^{j_1}e_{j_1}+\tilde{w}^{j_1}e_{j_0}\in \mathcal{SU}^+(\mathcal{R}^n,d_e)$ where $e_k$ is an n-dimensional vector with the $k$-th component being one and the other components being zero, we can get that
\begin{equation}
\eqalign{
\left<\hat{w},p\right>&=\left<\tilde{w}-\tilde{w}^{j_1}e_{j_1}+\tilde{w}^{j_1}e_{j_0},p\right>  \\
&=\left<\tilde{w},p\right>+\tilde{w}^{j_1}\left<e_{j_0},p\right>-\tilde{w}^{j_1}\left<e_{j_1},p\right>   \\
&=\left<\tilde{w},p\right>+\tilde{w}^{j_1}(p^{j_0}-p^{j_1})  \\
&\geq \left<\tilde{w},p\right>,}
\end{equation}
which conflicts with the initial hypothesis that $\tilde{w}$ maximizes $\left<w,p\right>$.

Denote the truncation of $p$ with respect to the index set $I$ as $\hat{p}=\sum_{k\in I} p^k e_k$. According to the equation (\ref{optimal_w}),  $\tilde{w} = \hat{p}/\|\hat{p}\|_2\in \mathcal{SU}^+(\mathcal{R}^n,d_e)$ maximizes $\left<w,p\right>$. On the other hand, $\left<\tilde{w},n\right>=\frac{1}{\|\hat{p}\|_2}\left<\hat{p},n\right>=0$, which means $\tilde{w}$ maximizes $\left<w,n\right>$ at the same time.
Consequently, $\tilde{w}=\hat{p}/\|\hat{p}\|_2$ is the maxima of $\left<w,p\right>+\left<w,n\right>$ and $\Delta Q^v$.
In order to facilitate further discussion and explanation, we define the projection function which maps $p$ to the corresponding maxima as
\begin{equation}
\mathcal{P}_{\mathcal{SU}^+(\mathcal{R}^n,d_e)}(p) = \hat{p}/\|\hat{p}\|_2.
\end{equation}

\subsection{Vector-label Propagation Algorithm(VLPA)}
\begin{figure*}
	\centering
	\includegraphics[width=0.9\textwidth]{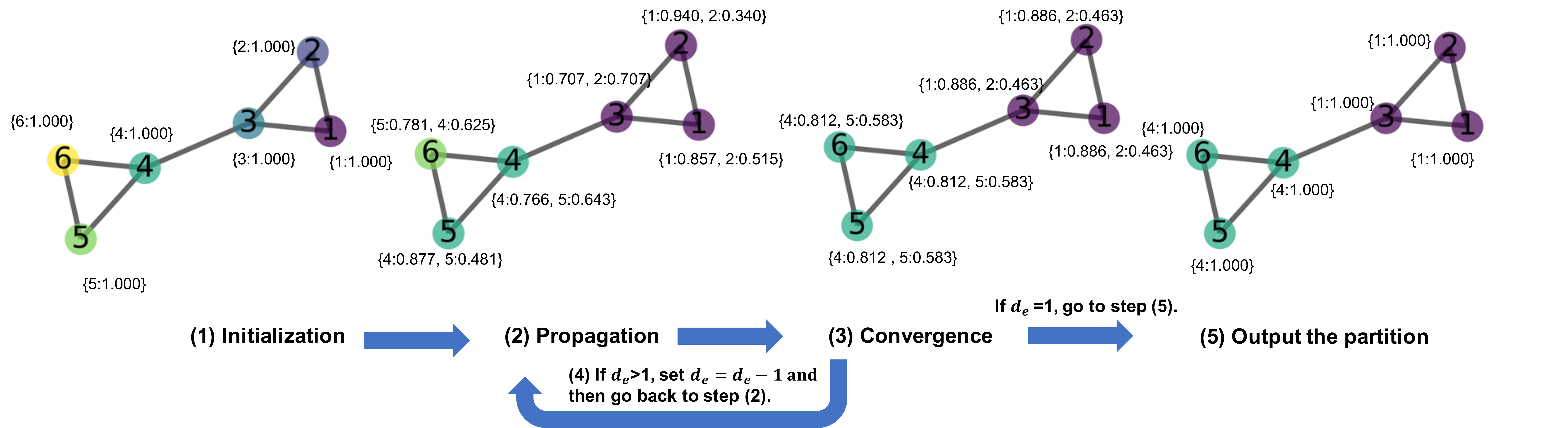}
	\caption{Visualization of the steps of VLPA.
		Each round is made of two steps: The first step is to perform vector-label propagation until convergence; In the second step, $d_e$ decreases by 1. The rounds are repeated iteratively until $d_e = 1$ and then output the community partition.}
	\label{diagram}
\end{figure*}

Based on the previous derivation, the details of the vector-label propagation algorithm(VLPA) are described as the following steps.
The schematic diagram of VLPA is shown in Fig \ref{diagram}.
\begin{enumerate}
	\item For each node $i$, initialize vector-label $v_i= e_i$ and $d_e$.
	\item For each node $i$, calculate $p_i$ and update vector-label $v_i$ by the projection function:
	$$v_i= \mathcal{P}_{\mathcal{SU}^+(\mathcal{R}^n, d_e)}(p_i).$$
	
	\item Repeat step 2  until convergence and no more than $M$ times.
	\item If $d_e>1$, set $d_e = d_e -1$ and then go back to step 2. Else, go to step 5.
	\item Output the partition that the community label of node $i$ is
  $$c_i=\arg\max_{k} \left<v_i,e_k\right>.$$
\end{enumerate}
Here are some explanations for the algorithm. First, $d_e$ is initialized as an integer greater than one so that some non-dominant labels are able to propagate farther and vector-labels can get closer to the global maxima.
Second, the parameter $M$ guarantees that the algorithm will stop.
Thirdly, vector-labels are converted to a normal partition in step 5. Finally, VLPA is convergent since each update is locally optimal and VLPA degenerates to Louvain when $d_e = 1$.

\subsection{Stochastic Vector-label Propagation Algorithm(sVLPA)}
In this part, we introduce the details of our second vector-label propagation algorithm sVLPA.
The stochastic gradient method is usually employed to avoid the local optima and increase the speed of algorithms in optimization problems.
Therefore, a stochastic gradient version of VLPA is taken into consideration to improve the performance furthur.
In practice, two strategies are proposed to provide randomness in the label propagation steps.
The first strategy is to select the essential dimension of each node from $\{1,2,\cdots,d_e\}$ uniformly and randomly, where the random essential dimension is denoted as $\mathbf{d_e}$.
The second strategy is to replace the truncated vector $\hat{p}_i$ with a random sparse vector $\mathbf{\hat{p}}_i$.
Referring to the previous derivation, the key step to calculate the updated vector-label is to obtain the index set $\{i_1,\cdots,i_{d_e}\}$ and $\hat{p}_i$.
For the purpose of increasing the randomness to prevent the local optima, we let $\mathbf{X}$ be a random variable selected from $\{1,\cdots,n\}$ with the probability
\begin{equation}
P(\mathbf{X} = k)= \left<p_i, e_k \right>^2 = p_i^2(k).
\end{equation}
Then we sample $X_1,\cdots,X_{\mathbf{d_e}}$ independently from the distribution of $\mathbf{X}$ to obtain a random index set $\mathbf{I}=\{X_1,\cdots,X_{\mathbf{d_e}}\}$ and let $\mathbf{\hat{p}} _i = \sum_{k\in \mathbf{I}} p_i(k)e_i$.

In contrast to VLPA where the indexes with the largest components are selected, sVLPA chooses each coordinate index $k$ randomly with a probability proportional to $p^2_i(k)$, which ensures that the index with a larger component is retained with a higher probability and increases the randomness of label updates.
The  vector-label after stochastic update can be represented by $\mathbf{\hat{p}} _i/\|\mathbf{\hat{p}}_i\|_2$.
The proposed sVLPA is described as below:
\begin{enumerate}
	\item For each node $i$, initialize vector-label $v_i= e_i$ and $d_e$.
	\item For each node $i$, calculate $p_i$, $\mathbf{d_e}$ and $\mathbf{\hat{p}}_i$ as we described above. Afterwards update the vector-label by
	    $$v_i= \mathbf{\hat{p}} _i /\|\mathbf{\hat{p}}_i\|_2.$$
	\item Repeat step 2 until convergence and no more than $M$ times.
	\item For each node $i$, calculate the determined $p_i$ and update vector-label $v_i$ by the projection function:
	$$v_i= \mathcal{P}_{\mathcal{SU}^+(\mathcal{R}^n, d_e)}(p_i).$$
	\item Repeat step 4 until convergence and no more than $M$ times.
	\item If $d_e>1$, set $d_e=d_e-1$ and go back to step 4. Else go to step 7.
	\item Output the partition that the community label of node $i$ is
  $$c_i=\arg\max_{k} \left<v_i,e_k\right>.$$
\end{enumerate}
As the only difference between sVLPA and VLPA is that the first round of label propagation(step 2) is stochastic in sVLPA, sVLPA will converge to a maximum just like VLPA.

\section{Experimental Results}\label{experiments}
In order to verify the validity of VLPA and sVLPA, we compare them with three popular community detection algorithms including Infomap \cite{rosvall-2008-maps-random-walks}, Louvain \cite{blondel-2008-fast-unfol-commun} and LPA \cite{raghavan-2007-near-linear-time} on both artificial and real networks.
\subsection{Setting of key parameters}
 Two key parameters, the essential dimension $d_e$ and the maximal number of iterations $M$, need to be determined before VLPA and sVLPA are applied to the networks.

To determine the optimal value of $d_e$, an LFR network is used to test VLPA with mixing parameter $\mu=0.6$, average degree $\left<k\right>=15$ and the number of nodes $n=1000$. Considering that the number of communities each node belongs to does not exceed the number of its neighbors, $d_e$ is set in the test as 15, the average degree of the network.
We pause VLPA at the end of step 3 and collect the maximal cut-off norm of vector-labels with different cut-off dimension $d$, which is the square root of the quadratic sum of the $d$ largest components of the vector-label $v$.

\begin{figure}
	\centering
	\includegraphics[width=0.65\textwidth]{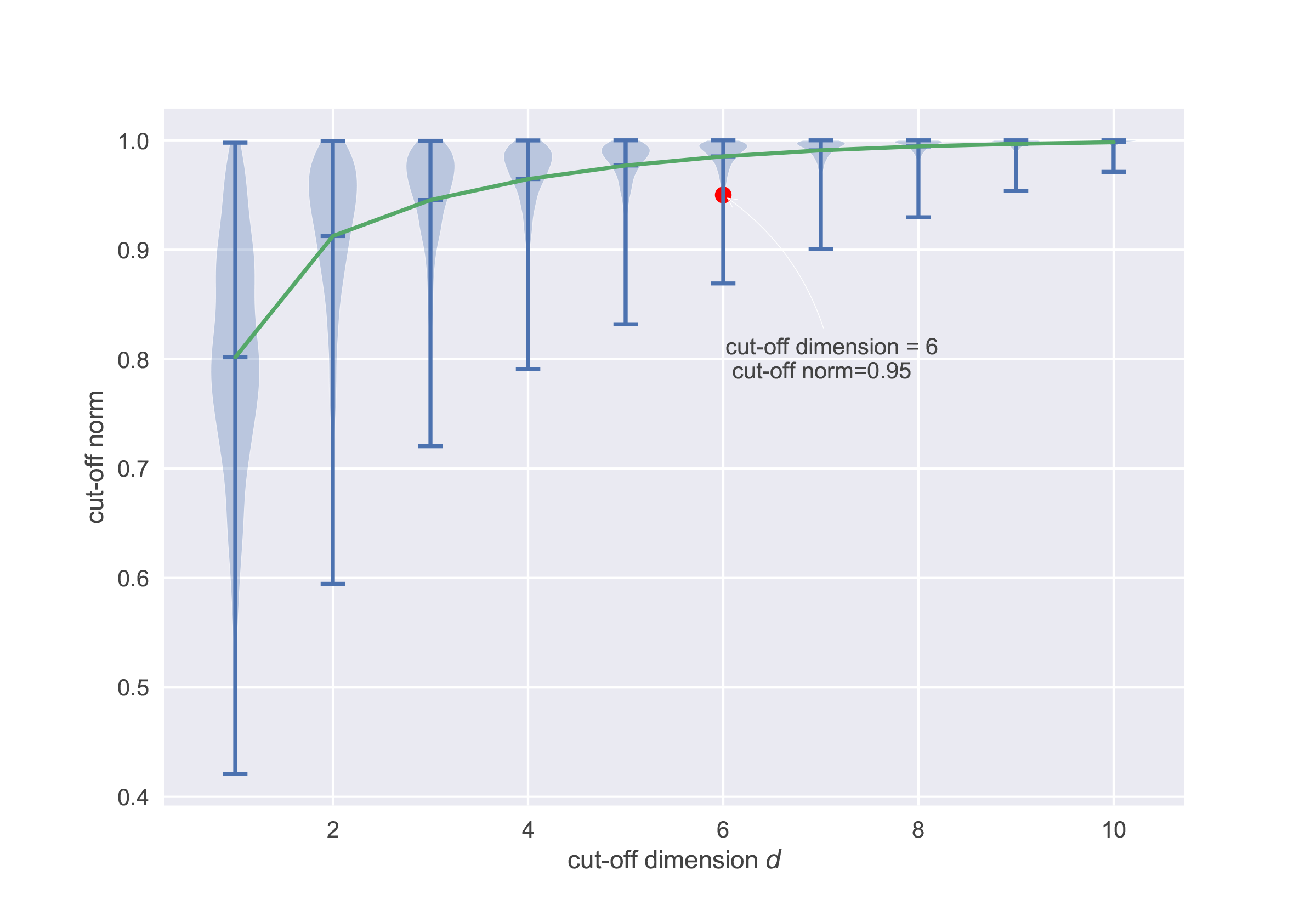}
	\caption{Distributions of the maximal cut-off norms of vector-labels under different cut-off dimensions.
		The LFR network is generated with mixing parameter $\mu=0.6$, average degree $\left<k\right>=15$, exponent parameters $\tau_1=\tau_2=2.0$ and  network size $n=1000$. Essential dimension $d_e$ is set to be $15$ during the process. The violin body provides the distribution under the corresponding cut-off dimension, the green line presents the mean value of the maximal cut-off norms and the coordinate of the marked red point is $(6,0.95)$.}
	\label{useful_dimension}
\end{figure}

Fig \ref{useful_dimension} exhibits the distributions of the maximal cut-off norms under different cut-off dimension $d$ at the end of step 3 in VLPA.
As the cut-off dimension $d$ increases, the average maximal cut-off norm increases and converges to 1.0. Meanwhile the distribution becomes more concentrated.
The marked red point indicates that when the cut-off dimension is 6, most vector-labels' maximal cut-off norms exceed $0.95$.
In other words, the norms of most vector-labels are concentrated within the six largest components, which inspires us to make $d_e$ less than or equal to 6 without losing too much vector-label information.

\begin{figure}
	\centering
	\includegraphics[width=0.65\textwidth]{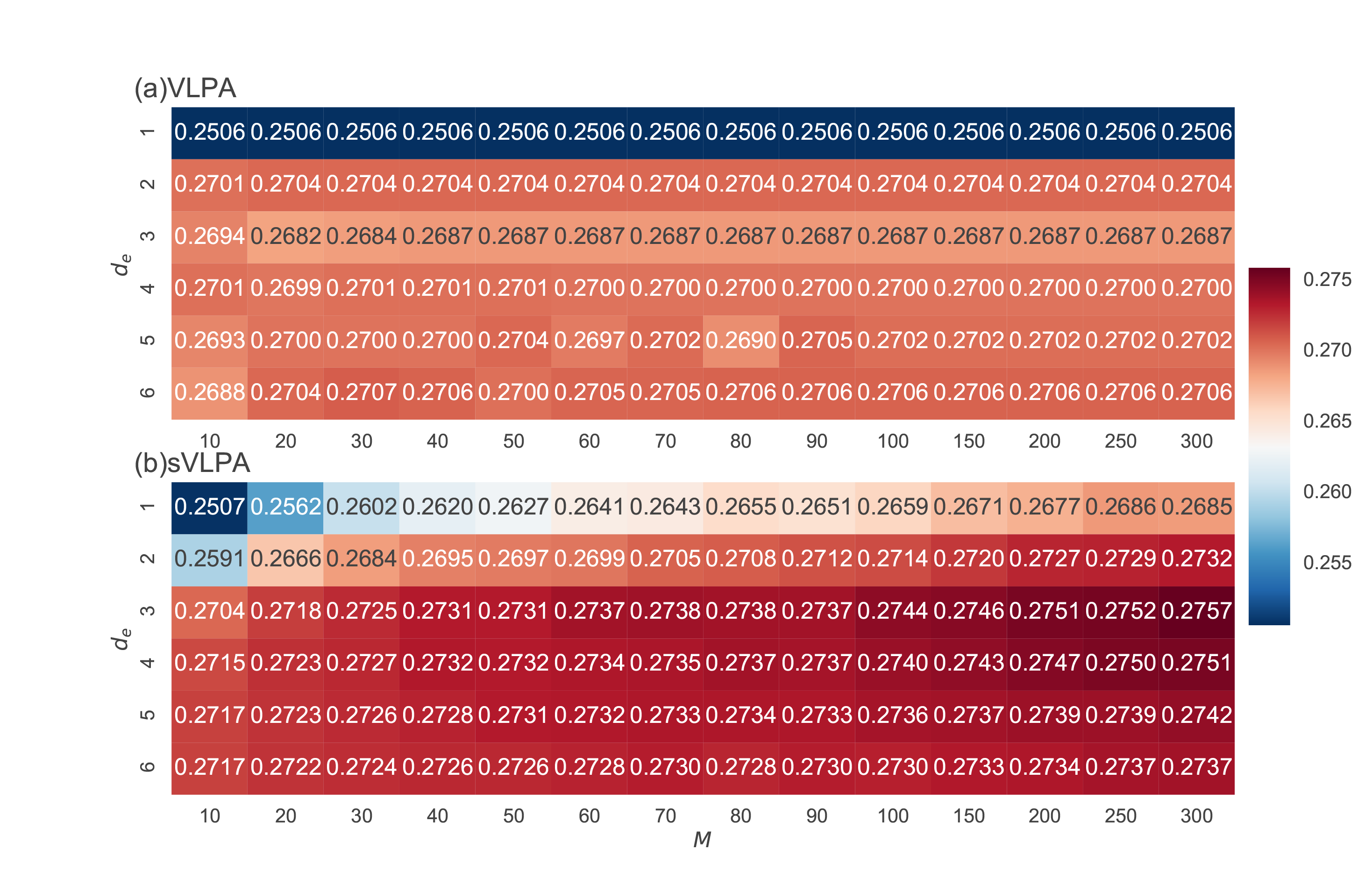}
	\caption{Heatmap of modularity optimized with different essential dimension $d_e$ and maximal iteration number $M$ on LFR benchmark networks. LFR networks are generated with $\mu=0.7$, $\left<k\right>=15$, $\tau_1=\tau_2=2.0$ and $n=1000$. The upper subgraph is the result of VLPA while the lower corresponds to the results of sVLPA. The color of each grid represents the average modularity of 100 trails.}
	\label{de_stepsize}
\end{figure}

In addition to $d_e$, another important parameter is the maximal number of iterations $M$ in each round.
We test the two algorithms to optimize the modularity with different $d_e$ and $M$ on an LFR network.
Figure \ref{de_stepsize} is a heat map of the modularity optimized by the algorithms where the upper subgraph is the result of VLPA and the lower subgraph is the result of sVLPA.

Compared with $d_e=1$, the optimization results of the algorithms are significantly improved when $d_e>1$, which means that the vector-label does have specific practical significance.
 As $d_e$ increases, the modularity initially increases and then declines.
 The reason for this phenomenon may be that when $d_e$ is small, $Q^v$ is approximate to $Q$ resulting in the higher modualrity. However, $Q^v$ is quite different from Q as $d_e$ grows larger. Therefore the results become worse.
Moreover, the optimal $d_e$ remains the same for large iterations $M$, indicating that the selection of the fixed parameter $d_e$ is relatively reliable.
Although the results of VLPA show that the modularity is slightly higher as $d_e=6$, $d_e=2$ and $M=20$ are selected in VLPA considering the time cost; For sVLPA, $d_e =3$ and $ M = 100$ are chosen.
All subsequent experimental results are based on the above parameter setting.

\subsection{Results on artificial networks}\label{artificial}
In the experiment, VLPA and sVLPA are compared with three aforementioned methods on LFR benchmark networks \cite{lancichinetti-2008-bench-graph-testin}.
Normalized mutual information(NMI) \cite{danon-2005-compar-commun-struc-ident}, as a widely used measure, is applied to evaluate the similarity between the predicted partition and the ground truth.

\begin{figure}
\centering
\includegraphics[width=0.65\textwidth]{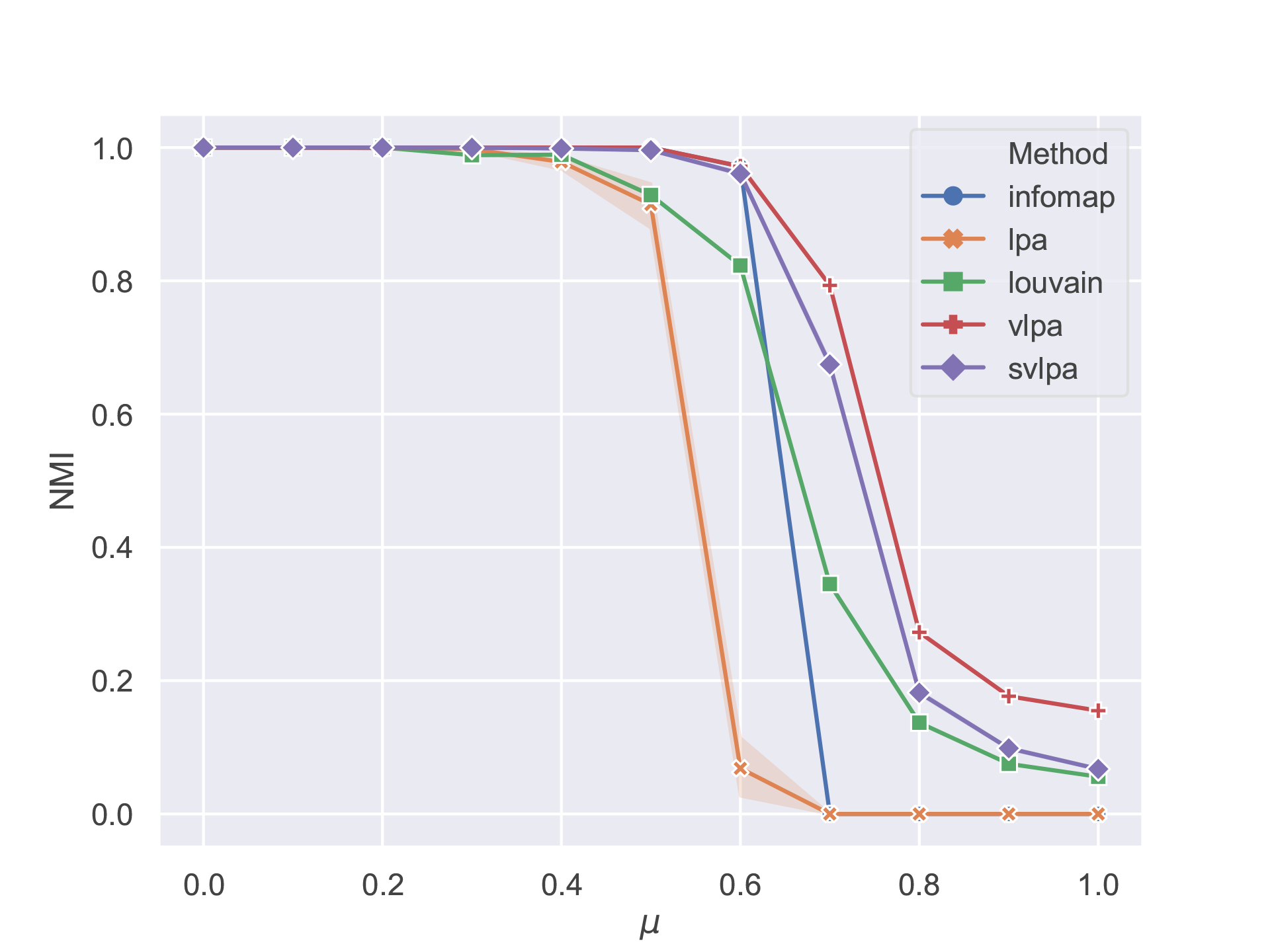}
\caption{Comparison on LFR benchmark networks by NMI as a function of the mixing parameter $\mu$.
LFR networks are generated with $n=1000$, $\tau_1=\tau_2 =2.0$ and $\left<k \right>=15$.
Each data point is the average of 100 trails while the error band means the 95\% confidence interval. }\label{NMI_compare}
\end{figure}

As Fig \ref{NMI_compare} shows, VLPA and sVLPA are as good as the best comparison algorithm as $\mu$ is less than 0.7 while VLPA is better than all other methods as $\mu$ is greater than 0.6.
The results show that all of our algorithms have better NMI performance than other comparison algorithms.
Moreover, VLPA is able to detect the communities that are closer to the ground truth when community structures are weak.
As a consequence, VLPA and sVLPA can excavate more community information from LFR benchmark networks.

\begin{figure}[h]
	\centering
	\includegraphics[width=0.9\textwidth]{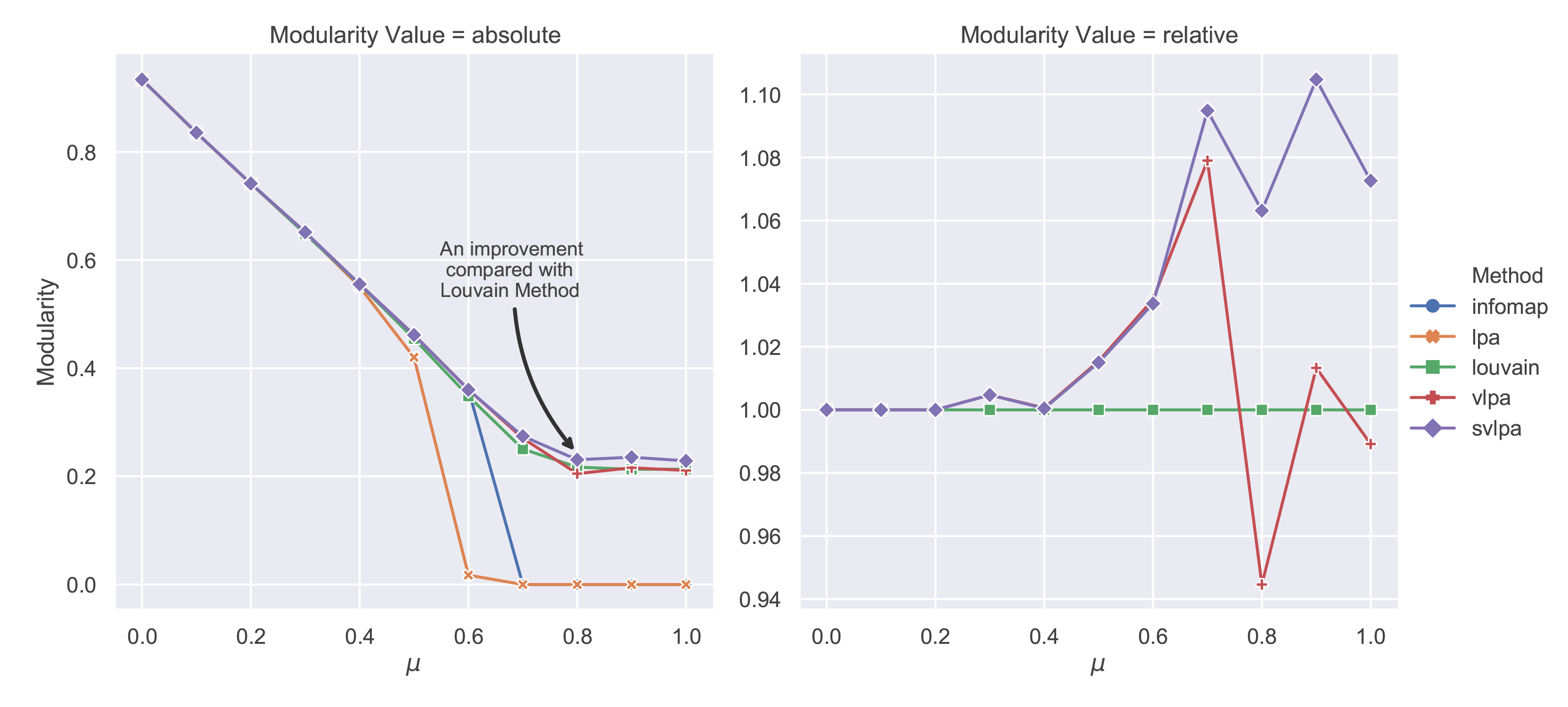}
	\caption{Modularity optimized by different methods as a function of the mixing parameter $\mu$. Panel (a) shows the absolute modualrity while panel (b) shows the relative modualrity compared to Louvain.
		LFR networks are generated with the same parameter configurations as Fig \ref{NMI_compare}.}
	\label{relative_modularity_compare}
\end{figure}

Since our methods are based on modularity optimization, it is also necessary to verify if VLPA and sVLPA perform well on maximizing modularity.
As shown in Fig \ref{relative_modularity_compare}(a), modularity has a significant downward trend indicating a declining community strength with the growth of $\mu$.
All methods perform well when $\mu$ is not greater than 0.6 while sVLPA outperforms the others on modualrity optimization when $\mu$ exceeds 0.6.
Louvain method, which detects community structure by optimizing modularity, performs relatively well in comparison algorithms.
Taken the modularity value obtained by Louvain method as the baseline,
the relative modularity of VLPA and sVLPA is illustrated in Fig \ref{relative_modularity_compare} (b) where the relative modularity refers to the ratio of the modularity obtained by one method to that obtained by Louvain.
The relative modularity of $\mu \geq 0.7$ shows that the improvement of sVLPA relating to Louvain method is about $6\% \sim 10\%$.
The precise modularity values of different methods are presented in Table \ref{mod_table}.
VLPA only loses to Louvain when $\mu$ is 0.8 while sVLPA is better than Louvain in all conditions.
For most mixing parameters $\mu$, the standard deviation of sVLPA is less than 1\%, which means that our algorithm is relatively stable.
Surprisingly, the maximal modularity of $\mu=1.0$ may indicate that even LFR networks are randomly generated without any underlying community structure, the randomness of networks may provide some weak community structures just as $\mu=0.8$.

\begin{table*}
\centering
\resizebox{\textwidth}{!}{
		\begin{tabular}{llllllll}
\hline
\hline
			
			$\mu$ & benchmark &            Infomap &            Louvain &             LPA &                 sVLPA &               vLPA & relative \\
			\hline
			0.0 &    0.9338 &  \redb{0.9338}$\pm$0.0 &  \redb{0.9338}$\pm$0.0 &  0.9336$\pm$0.0013 &     \redb{0.9338}$\pm$0.0 &  \redb{0.9338}$\pm$0.0 &     0.0\% \\
			0.1 &    0.8355 &  \redb{0.8355}$\pm$0.0 &  \redb{0.8355}$\pm$0.0 &  0.8353$\pm$0.0014 &     \redb{0.8355}$\pm$0.0 &  \redb{0.8355}$\pm$0.0 &     0.0\% \\
			0.2 &    0.7416 &  \redb{0.7416}$\pm$0.0 &  \redb{0.7416}$\pm$0.0 &   0.741$\pm$0.0023 &  \redb{0.7416}$\pm$0.0001 &  \redb{0.7416}$\pm$0.0 &     0.0\% \\
			0.3 &    0.6515 &  \redb{0.6515}$\pm$0.0 &        0.6484$\pm$0.0 &  0.6506$\pm$0.0017 &     \redb{0.6515}$\pm$0.0 &  \redb{0.6515}$\pm$0.0 &   0.478\% \\
			0.4 &    0.5554 &  \redb{0.5554}$\pm$0.0 &         0.555$\pm$0.0 &  0.5507$\pm$0.0067 &        0.5553$\pm$0.0004 &  \redb{0.5554}$\pm$0.0 &   0.054\% \\
			0.5 &    0.4617 &  \redb{0.4617}$\pm$0.0 &        0.4546$\pm$0.0 &  0.4202$\pm$0.1086 &        0.4614$\pm$0.0004 &  \redb{0.4617}$\pm$0.0 &   1.496\% \\
			0.6 &    0.3601 &  \redb{0.3607}$\pm$0.0 &        0.3483$\pm$0.0 &  0.0174$\pm$0.0764 &        0.3601$\pm$0.0008 &        0.3605$\pm$0.0 &   3.388\% \\
			0.7 &    0.2675 &           0.0$\pm$0.0 &        0.2506$\pm$0.0 &        0.0$\pm$0.0 &  \redb{0.2743}$\pm$0.0012 &        0.2704$\pm$0.0 &   9.457\% \\
			0.8 &    0.1665 &           0.0$\pm$0.0 &        0.2171$\pm$0.0 &        0.0$\pm$0.0 &  \redb{0.2308}$\pm$0.0019 &        0.2051$\pm$0.0 &    6.31\% \\
			0.9 &    0.0674 &           0.0$\pm$0.0 &        0.2131$\pm$0.0 &        0.0$\pm$0.0 &  \redb{0.2354}$\pm$0.0022 &        0.2159$\pm$0.0 &  10.465\% \\
			1.0 &    -0.032 &           0.0$\pm$0.0 &        0.2133$\pm$0.0 &        0.0$\pm$0.0 &  \redb{0.2288}$\pm$0.0024 &         0.211$\pm$0.0 &   7.267\% \\
			\hline
\hline
		\end{tabular}}
		
		\caption{Modularity optimized by different methods on LFR networks.
			The ``benchmark" column means the modularity of the ground truths while the ``relative" column is the relative increment of sVLPA over Louvain, expressed as a percentage.
			\redb{Red and bold font} indicates the best modularity among all the algorithms.
			The values in front of and behind $\pm$ are the average of 100 trials and the corresponding standard deviation.
			LFR networks are generated with the same parameter configurations as Fig \ref{NMI_compare}.
		}
		\label{mod_table}
\end{table*}	

In 2009,  Lambiotte {\it et al.} proposed a modified modularity with resolution parameters $t$ \cite{lambiotte2009laplacian}.
Using the previously defined symbols, the modified modularity is denoted as
\begin{equation}
Q(t)=(1-t)+\frac{1}{2m}\sum_{i j}\left(tA_{i j}-\frac{k_i k_j}{2m}\right)\delta (c_i, c_j).
\end{equation}
Obviously, we can generalize this modified modularity by applying the vector-label as
\begin{equation}
Q^v(t) = (1-t)+\frac{1}{2m}\sum_{i j}\left(tA_{i j}-\frac{k_i k_j}{2m}\right) \left <v_i, v_j \right> .
\end{equation}
When $t=1$, the modified modularity degenerates into conventional modularity.
For modified modularity with different resolution parameters(time $t$), similar results are acquired and displayed in Tab \ref{modified_mod_table}.
Obviously sVLPA performs better than Louvain and VLPA.
The above experiments of modified modularity optimization means that our vector-label framework can be applied to most community detection algorithms based on partition metrics that can be represented by labels and $\delta$ function.

\begin{table}
\centering
\setlength{\tabcolsep}{7mm}{
\begin{tabular}{llll}
\hline
\hline
	Resolution &         Louvain &                 sVLPA &          vLPA \\
	\hline
	0.2 &  0.8285$\pm$0.0004 &   \redb{0.831}$\pm$0.0035 &   0.8293$\pm$0.0 \\
	0.4 &  0.6849$\pm$0.0016 &  \redb{0.6897}$\pm$0.0001 &   0.6863$\pm$0.0 \\
	0.6 &  0.5401$\pm$0.0029 &  \redb{0.5497}$\pm$0.0003 &   0.5468$\pm$0.0 \\
	0.8 &  0.3979$\pm$0.0037 &  \redb{0.4105}$\pm$0.0008 &   0.4086$\pm$0.0 \\
	1.0 &   0.2606$\pm$0.004 &  \redb{0.2758}$\pm$0.0012 &   0.2704$\pm$0.0 \\
	1.2 &   0.1336$\pm$0.004 &   \redb{0.1491}$\pm$0.002 &   0.1332$\pm$0.0 \\
	1.4 &  0.0298$\pm$0.0057 &  \redb{0.0428}$\pm$0.0045 &  -0.0044$\pm$0.0 \\
\hline
\hline
\end{tabular}
\caption{Modified modularity optimized with different resolution parameters $t$ on LFR networks.
\redb{Red and bold font} indicates the best modified modularity obtained among VLPA, sVLPA and Louvain.
The value in front of and behind $\pm$ is the average of 100 trials and the corresponding standard deviation.
LFR networks are generated by adopting the same configuration parameters as Fig \ref{de_stepsize}.
}}
\label{modified_mod_table}
\end{table}

\begin{table*}[!ht]\tiny
\centering
\resizebox{\textwidth}{!}{
\begin{tabular}{lllllllll}
\hline
\hline
	
	Network &          n &           m &              Infomap &               Louvain &               LPA &                   sVLPA &                vLPA &              relative \\
	\hline
	karate\cite{zachary1977an} &         34 &          78 &        0.402(3.18ms) &        0.419(0.131ms) &   0.341(0.0853ms) &           0.415(2.28ms) &  \redb{0.42}(0.41ms) &          -0.955\%(17.4) \\
	dolphins\cite{lusseau2003bottlenose} &         62 &         159 &  \redb{0.523}(6.91ms) &        0.519(0.329ms) &    0.466(0.175ms) &      \redb{0.523}(5.46ms) &        0.5(0.934ms) &  \textbf{0.771\%}(16.6) \\
	football\cite{girvan-2002-commun-struc-social} &        115 &         613 &        0.601(9.58ms) &  \redb{0.604}(0.562ms) &    0.586(0.262ms) &           \redb{0.604}(7.34ms) &       0.603(1.96ms) &    \textbf{0.0\%}(13.1) \\
	email-Eu-core\cite{leskovec-2007-graph, yin-2017-local}&       1005 &       16385 &        0.061(0.551s) &  \redb{0.439}(0.0159s) &     0.002(5.26ms) &           0.434(0.119s) &      0.428(0.0543s) &          -1.139\%(7.52) \\
	CollegeMsg\cite{panzarasa-2009-patterns} &       1899 &       13838 &         0.052(1.22s) &        0.254(0.0329s) &      0.001(4.9ms) &     \redb{0.275}(0.152s) &      0.259(0.0658s) &  \textbf{8.268\%}(4.61) \\
	soc-sign-bitcoinalpha\cite{kumar-2016-edge, kumar-2018-rev2}  &       3683 &       12972 &         0.403(3.49s) &         0.49(0.0239s) &    0.112(0.0137s) &       \redb{0.5}(0.156s) &      0.486(0.0654s) &  \textbf{2.041\%}(6.53) \\
	facebook\_combined\cite{leskovec2012learning}&       4039 &       88234 &          0.81(2.48s) &  \redb{0.835}(0.0523s) &     0.809(0.028s) &           0.832(0.276s) &        0.831(0.11s) &          -0.359\%(5.28) \\
	ca-GrQc\cite{leskovec-2007-graph}  &       5242 &       14490 &        0.234(0.503s) &  \redb{0.863}(0.0311s) &    0.793(0.0359s) &           0.854(0.155s) &      0.825(0.0625s) &          -1.043\%(4.98) \\
	soc-sign-bitcoinotc\cite{kumar-2016-edge,kumar-2018-rev2} &       5573 &       18591 &         0.406(6.36s) &         0.48(0.0246s) &    0.116(0.0229s) &      \redb{0.521}(0.23s) &      0.509(0.0916s) &  \textbf{8.542\%}(9.36) \\
	p2p-Gnutella08\cite{ripeanu-2002-mapping,leskovec-2007-graph}  &       6301 &       20777 &          0.41(9.88s) &  \redb{0.464}(0.0717s) &    0.012(0.0214s) &           0.461(0.319s) &       0.412(0.117s) &          -0.647\%(4.45) \\
	wiki-Vote\cite{leskovec-2010-signed, leskovec-2010-predicting}&       7115 &  1.0076$\times 10^5$ &         0.064(11.3s) &         0.418(0.063s) &    0.035(0.0386s) &     \redb{0.429}(0.614s) &         0.42(0.29s) &  \textbf{2.632\%}(9.76) \\
	ca-HepPh  \cite{leskovec-2007-graph}&  1.201$\times 10^4$ &  1.1850$\times 10^5$ &         0.062(3.93s) &         0.655(0.191s) &     0.454(0.183s) &     \redb{0.656}(0.908s) &       0.645(0.396s) &  \textbf{0.153\%}(4.75) \\
	ca-AstroPh\cite{leskovec-2007-graph} &  1.877$\times 10^4$ &  1.9808$\times 10^5$ &         0.057(7.59s) &         0.615(0.331s) &     0.296(0.308s) &      \redb{0.621}(1.97s) &        0.604(0.82s) &  \textbf{0.976\%}(5.96) \\
	ca-CondMat\cite{leskovec-2007-graph}&  2.313$\times 10^4$ &       93468 &         0.155(3.97s) &   \redb{0.728}(0.213s) &     0.616(0.404s) &            0.717(1.16s) &        0.67(0.471s) &          -1.511\%(5.45) \\
	email-Enron\cite{leskovec-2009-community,klimt2004introducing} &  3.669$\times 10^4$ &  1.8383$\times 10^5$ &          0.129(9.1s) &          0.586(0.32s) &     0.309(0.328s) &      \redb{0.626}(1.81s) &         0.62(0.73s) &  \textbf{6.826\%}(5.64) \\
	email-EuAll\cite{leskovec-2007-graph}&  2.652$\times 10^5$ &  3.6502$\times 10^5$ &          0.63(40.1s) &         0.768(0.478s) &      0.689(1.93s) &       \redb{0.78}(8.05s) &        0.766(3.06s) &  \textbf{1.562\%}(16.9) \\
	com-dblp.ungraph\cite{yang2015defining}&  3.171$\times 10^5$ &  1.0499$\times 10^6$ &             NaN &    \redb{0.809}(2.37s) &      0.688(47.4s) &            0.797(18.6s) &        0.752(7.92s) &          -1.483\%(7.82) \\
	com-amazon.ungraph\cite{yang2015defining}&  3.349$\times 10^5$ &  9.2587$\times 10^5$ &             NaN &    \redb{0.926}(2.27s) &      0.785(25.8s) &            0.887(19.8s) &         0.843(8.4s) &          -4.212\%(8.72) \\
	com-youtube.ungraph\cite{yang2015defining}&  1.135$\times 10^6$ &  2.9876$\times 10^6$ &             NaN &          0.685(7.39s) &       0.626(113s) &        \redb{0.725}(68s) &        0.703(28.2s) &  \textbf{5.839\%}(9.21) \\
	soc-pokec-relationships\cite{takac2012data}&  1.633$\times 10^6$ &  2.2302$\times 10^7$ &             NaN &           0.696(147s) &       0.696(417s) &       \redb{0.724}(512s) &         0.721(220s) &  \textbf{4.023\%}(3.49) \\
	wiki-topcats\cite{yin2017local,klymko2014using} &  1.791$\times 10^6$ &  2.5446$\times 10^7$ &             NaN &           0.653(133s) &       0.207(241s) &       \redb{0.659}(528s) &         0.655(241s) &  \textbf{0.919\%}(3.95) \\
	wiki-Talk\cite{leskovec-2010-signed,leskovec-2010-predicting}&  2.394$\times 10^6$ &  4.6596$\times 10^6$ &             NaN &    \redb{0.566}(10.4s) &        0.45(158s) &             0.564(222s) &        0.559(91.9s) &          -0.353\%(21.4) \\
	soc-LiveJournal1\cite{backstrom2006group,leskovec-2009-community} &  4.848$\times 10^6$ &   4.311$\times 10^7$ &             NaN &           0.737(547s) &  0.547(1.01$\times 10^3$s) &  \redb{0.761}(1.15$\times 10^3$s) &         0.745(481s) &  \textbf{3.256\%}(2.11) \\
\hline
	\hline
\end{tabular}}
\caption{Modularity optimization results of different methods applied to real networks.
\redb{Red and bold font} indicates the best modularity among all the algorithms while the number in the parentheses is the running time of the algorithm.
The ``relative" column shows the relative increment of modularity(\textbf{Bold font} means positive) and the relative time of sVLPA to Louvain.
NaN means the running time is over 1000 seconds.
Each value is the average of 10 trials.
All experiments are conducted on the same machine(Macbook Pro 2017, 16G RAM).}
\label{real_networks}
\end{table*}

\begin{figure}[h]
	\centering
	\includegraphics[width=0.65\textwidth]{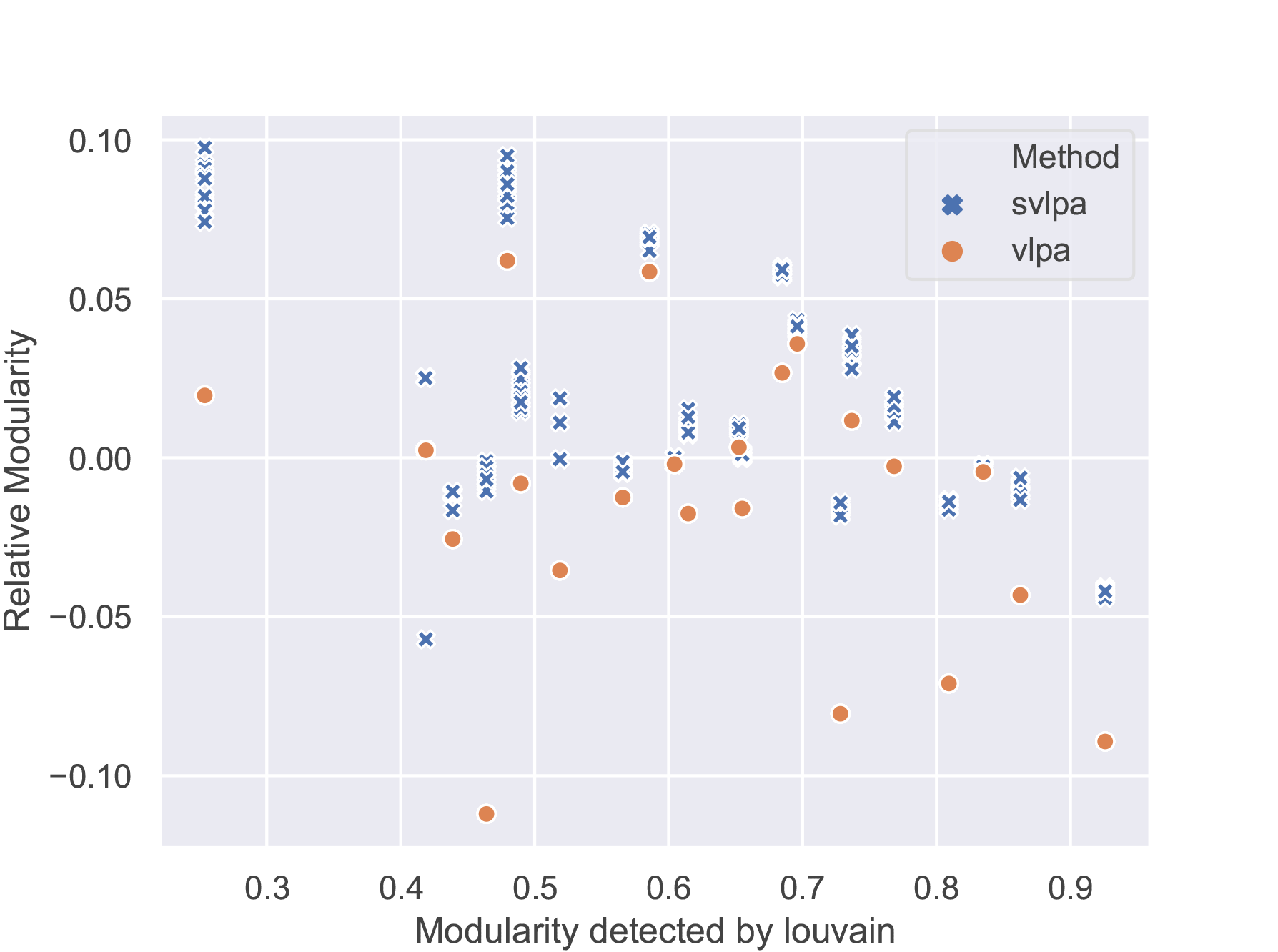}
	\caption{Results of relative modularity over Louvain on real networks.  Each point represents one experiment where the x-coordinate represents the modularity detected by Louvain and the y-coordinate represents the relative improvement in modularity by VLPA or sVLPA. }
	\label{relative_modularity_real}
\end{figure}

\subsection{Experiments on real networks}
Experimental results on real networks are presented in Tab \ref{real_networks} where most networks come from Stanford Large Network Dataset Collection (SNAP) \cite{snapnets} and the largest one contains more than 4 million nodes.
In Tab \ref{real_networks}, sVLPA performs best on more than half of the real networks, especially when the modularity value is relativity small.
Louvain is still better than sVLPA in some networks while the difference of modualrity is not apparent and usually less than 1\%. However sVLPA can improve by 4\% $\sim$ 8\% on most cases.
It is observed that the modularity with relatively high promotion is around 0.25$\sim$0.7, and when the modularity exceeds 0.7 sVLPA is more likely to lose to Louvain.
In Fig \ref{relative_modularity_real}, each point represent one experiment on a real network.
It is evident that our sVLPA has a better performance over Louvain in real networks with weaker community structures, which means that our algorithm is indeed able to jump out of the local maxima.

Experiments on synthetic and real networks suggest that sVLPA has the ability to jump out of the local maxima and get closer to the global maxima than Louvain, especially when the community structure is relativity weak.

\begin{figure}
	\centering
	\includegraphics[width=0.65\textwidth]{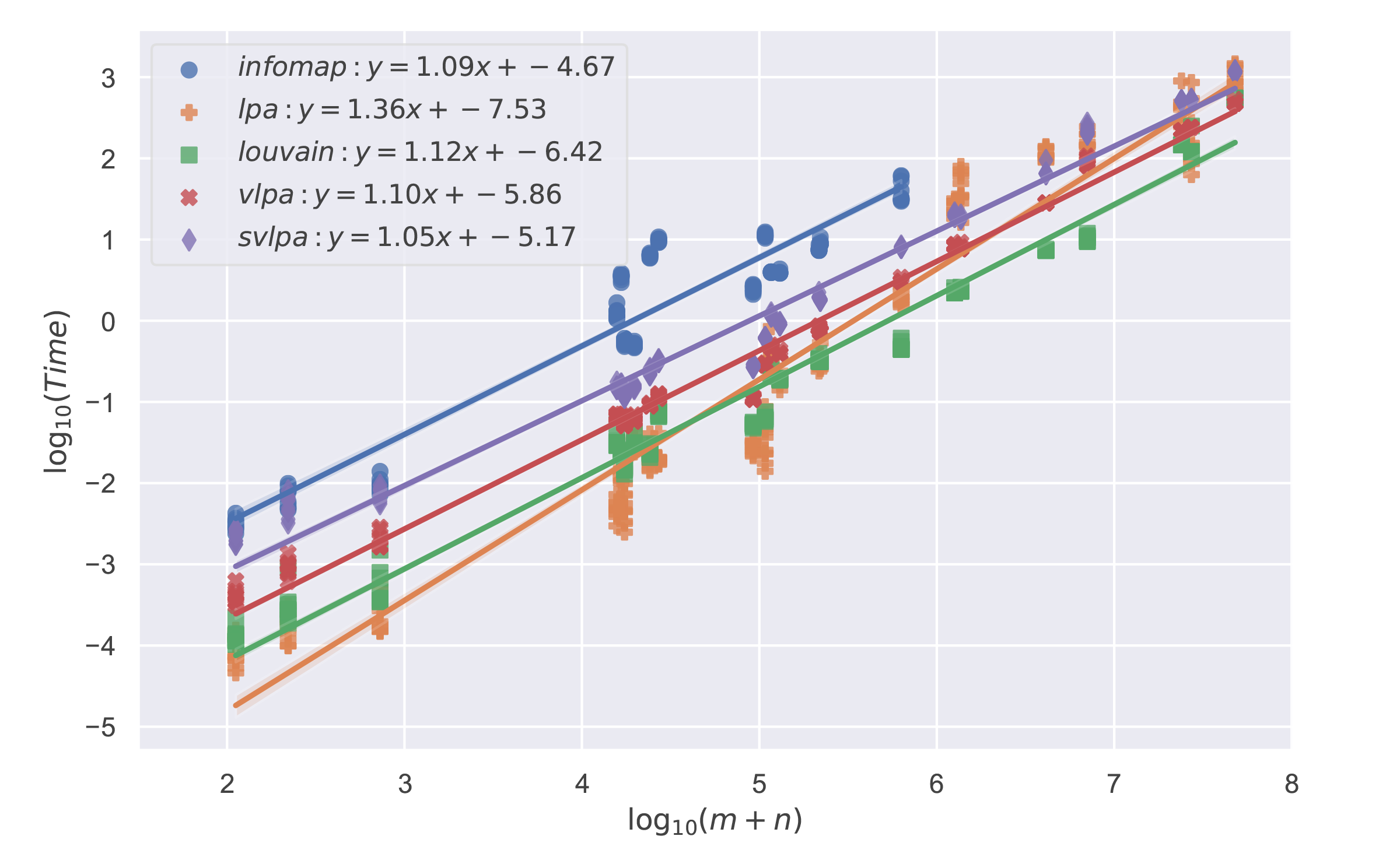}
	\caption{Running times of different methods on real networks.
		Each point represents one experiment where the x-coordinate is the logarithm of the sum of edge number and node number and the y-coordinate is the logarithm of the running time.
		The fitting lines are calculated by the least square method.
}
	\label{time_scale}
\end{figure}

As for the efficiency of algorithms, the relationships between the running time and the network scale are displayed in Fig \ref{time_scale}. Although the absolute running time of sVLPA is higher on most real networks, the experimental time complexity(the slope of the fitting line) is a little better than Louvain and LPA.
In the next section, we will provide the therotical analysis that sVLPA and VLPA have linear time complexity of $O(m+n)$.

\section{Time complexity}\label{time}
In this section, we provide the theoretical time complexity of our algorithms.
Let $n$ be the number of nodes, $m$ be the number of edges and $c$ be the number of iterations before convergence. The time cost of each step is analyzed below.
\begin{enumerate}
	\item Initialization requires $O(n)$ time.
	\item In each iteration, it takes $O(d_e (k_i+1))$ to calculate $p_i$  and $O(d_e k_i)$ to calculate $\hat{p}_i/\|\hat{p}_i\|_2$ or $\mathbf{\hat{p}}_i/\|\mathbf{\hat{p}}_i\|_2$ for node i. Therefore each iteration requires $O(d_e(m+n))$ time in total.
	\item Generating the final partition needs $O(n)$ time.
\end{enumerate}
Since the number of iterations $c$ is no more than the maximal iteration number $M$, the time complexity of VLPA as well as sVLPA is $O(M d_e (m+n))$ in total.

\section{Conclusions and Further Discussions}\label{conclusions}

Detecting the community structures of complex networks has always been a widely concerned problem in many different fields.
This paper has developed a novel framework that turnes the community detection problem into a continuous optimization problem by introducing the newly defined vector-label and the vector-modularity. Under such a scheme, two new algorithms are proposed: VLPA and sVLPA. Compared with the previous algorithms with traditional one-dimensional label, VLPA allows the propagation of a series of weighted labels, which prevents the inferior labels from disappearing too fast. Therefore, VLPA can utilize more topological information and obtain better performance when the community structure is relatively weak. Further, sVLPA incorporates randomness into the vector-label updating process via a stochastic gradient method, which greatly increases the probability of reaching the global optimum. Experimental results show that sVLPA outperforms Louvain on all LFR benchmarks and on most of the real-world networks. Moreover, both VLPA and sVLPA have a near-linear time complexity $O(Md_e (m+n))$, which is comparable to LPA and Louvain method.

As shown in results \ref{artificial}, our theoretical framework based on vector-label propagation can also be applied to optimize other partition criteria like the modified modularity, which may pave ways for the development of more effective algorithms. Our work may also inspire future community detection studies on multi-layer or multi-label networks.

\section{Acknowledgement}
This work is supported by Program of National Natural Science Foundation of China Grant No. 11871004 and No. 11922102 and National Key Research and Development Program of China Grant No. 2018AAA0101100.

\providecommand{\newblock}{}

\end{document}